\documentclass[iop]{emulateapj}
\usepackage{graphicx,epsfig,amssymb,amsmath,color,lineno}

\def\eg{{e.g.,~}}
\def\ie{{i.e.,~}}
\def\lat{{{\it Fermi}-LAT~}}

\begin{document}
\label{firstpage}

\title{Spectral analysis of {\it Fermi}-LAT blazars above 50~GeV}

\author{{\sc Alberto Dom\'inguez}\altaffilmark{1,2} and {\sc Marco Ajello}\altaffilmark{1}}

\slugcomment{Published; November 4, 2015}

\shorttitle{Spectral analysis of {\it Fermi}-LAT blazars above 50~GeV}
\shortauthors{Dom\'inguez \& Ajello}

\altaffiltext{1}{Department of Physics \& Astronomy, Clemson University, Clemson, SC 29634, USA}
\altaffiltext{2}{Grupo de Altas Energ\'ias, Universidad Complutense, E-28040 Madrid, Spain}

\email{alberto@clemson.edu, majello@clemson.edu}


\begin{abstract}
We present an analysis of the intrinsic (unattenuated by the extragalactic background light, EBL) power-law spectral indices of 128 extragalactic sources detected up to $z\sim 2$ with the {\it Fermi}-Large Area Telescope (LAT) at very high energies (VHEs, $E\geq 50$~GeV). The median of the intrinsic index distribution is 2.20 (versus 2.54 for the observed distribution). We also analyze the observed spectral breaks (\ie the difference between the VHE and high energy, HE, $100~{\rm MeV}\leq E\leq 300$~GeV, spectral indices). The \lat has now provided a large sample of sources detected both at VHE and HE with comparable exposure that allows us to test models of extragalactic $\gamma$-ray photon propagation. We find that our data are compatible with simulations that include intrinsic blazar curvature and EBL attenuation. There is also no evidence of evolution with redshift of the physics that drives the photon emission in high-frequency synchrotron peak (HSP) blazars. This makes HSP blazars excellent probes of the EBL.
\end{abstract}


\section{Introduction}
\label{sec:intro}
Star formation activity and super-massive black hole accretion produce over time the extragalactic background light (EBL). This is an evolving diffuse radiation field that permeates the Universe from ultraviolet to far-infrared wavelengths \citep[\eg][]{hauser01,dwek13}. The interaction between extragalactic very high energy (VHE, $E\geq 50$~GeV) $\gamma$-ray photons and the EBL leads to an attenuation of the source emission that is energy and redshift dependent \citep[\eg][]{stecker92}. In fact, this attenuation has been measured from observations of blazars by \citet{ebl12}, \citet{hess_ebl13}, \citet{dominguez13a}, and \citet{biteau15}.

In general, ground-based Imaging Atmospheric Cherenkov Telescopes (IACTs) report their VHE spectral observations with a power-law fit, $dN/dE\propto E^{-\Gamma_{\rm obs}}$, which is parameterized by an observed spectral index $\Gamma_{\rm obs}$. Typically, this fit is a good description of the data in the relatively narrow energy range within with IACTs detect sources. At lower energies ($100~{\rm MeV}\leq E\leq 300$~GeV, defined as the high energy, HE, range) the {\it Fermi}-Large Area Telescope (LAT) is sensitive to photons over more decades of energy, which opens the possibility of resolving spectral curvature \citep{3FGL}. Importantly for the present work, the improvements provided by the new Pass 8 event-level analysis, in addition to the increased data volume, allow the \lat to detect sources at VHEs leading to the 2FHL catalog \citep{2FHL}.


The redshift evolution of the observed spectral break (\ie the difference between the observed spectral index at VHE and HE) is proposed by \citet{stecker06} and \citet{stecker10} as a proxy of blazar physics as well as EBL information. These authors study a limited sample of 13 BL~Lacs at $z<0.22$ observed by both the \lat and IACTs. Later, \citet{sanchez13} extended the analysis by enlarging the sample to 23 sources and up to $z\sim 0.5$. Alternatively, \citet{essey12} propose the observed spectral break as a test of the hypothesis that there may be a significant secondary component at VHEs. This secondary component would be produced by the interaction of cosmic rays emitted by the source with the cosmic microwave background \citep{essey10}.

It is necessary to correct the VHE observations for EBL attenuation in order to correctly interpret blazar physics. The EBL corrected/unattenuated spectrum is known as the intrinsic spectrum (\ie the spectrum that we would observe if there were no EBL). In fact, an interesting property that encodes the blazar physics is given by what we call the intrinsic spectral break. The intrinsic spectral break is defined as the difference between the intrinsic spectral index at VHEs and the spectral index at HEs, and it has not been systematically studied yet.


In this letter, we present a study of the intrinsic spectral indices of a statistical sample of blazars with known redshifts up to $z\sim 2$ detected by the \lat at VHEs. Their observed and intrinsic spectral breaks are analyzed both in the context of blazar physics and EBL attenuation. Section~\ref{sec:data} describes our data sample in terms of the intrinsic indices, whereas Section~\ref{sec:results} shows and discusses the results on the evolution of the spectral breaks over redshift. Finally, a summary of our analysis is presented in Section~\ref{sec:summary}.

\newpage

\section{Description of the data sample}
\label{sec:data}
The \lat 2FHL catalog presents a total of 360 sources over the whole sky detected at energies greater than 50~GeV. This catalog provides observed spectral indices, $\Gamma_{\rm obs}^{\rm 2FHL}$, which characterize the VHE power-law spectra (since at these energies the LAT is not sensitive enough to detect spectral curvature) from 50~GeV to 2~TeV, redshifts, and synchrotron peak frequencies for 128 extragalactic sources. Using the synchrotron peak frequency, we divide the sample into 33 low synchrotron peak blazars (LSP, $\log_{10}(\nu/{\rm Hz})\leq 14$) of which 10 are FSRQs, 12 intermediate synchrotron peak blazars (ISP, $14<\log_{10}(\nu/{\rm Hz})< 15$), and 83 high synchrotron peak blazars (HSP, $\log_{10}(\nu/{\rm Hz})\geq 15$); therefore our sample contains mostly HSP rather than LSP and ISP blazars. To derive the intrinsic spectral indices for these 128 sources, we modify the 2FHL analysis pipeline, which is thoroughly described by \citet{2FHL}, to include the EBL effect in the source models and fits. The EBL effect is taken from the model by \citet{dominguez11a}, which is derived from multiwavelength galaxy observations. The data set and analysis parameters are exactly the same as in the 2FHL.\footnote{The intrinsic indices are provided for the \citet{dominguez11a} and \citet{gilmore12} Fiducial models in the 2FHL catalog at http://fermi.gsfc.nasa.gov/ssc/data/access/lat/2FHL/}

\begin{figure}
\includegraphics[width=\columnwidth]{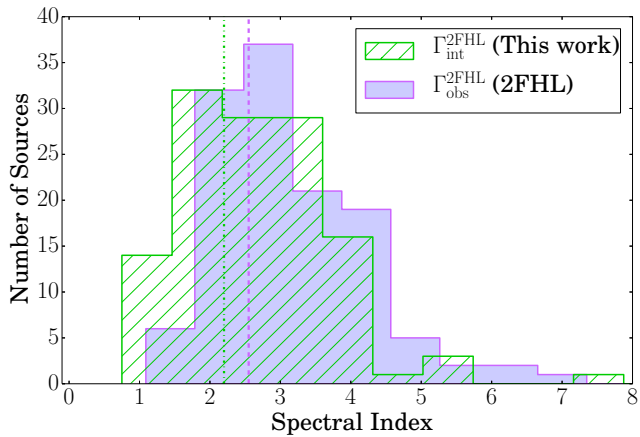}
\caption{Distribution of observed (purple) and intrinsic (green) spectral index at energies greater than 50~GeV. The mean values of the distributions, calculated using the methodology followed by \citet{venters07}, are shown with vertical lines.}
\label{fig1}
\end{figure}

Our analysis procedure leads to the histogram of spectral indices shown in Figure~\ref{fig1}, where the observed and intrinsic index distributions are compared. In order to estimate the mean and spread of these distributions (affected by large uncertainties in the individual measurements), we follow the likelihood methodology described by \citet{venters07}. As expected, the main effect of the EBL correction is to reduce the mean index (from 2.54 observed to 2.20 intrinsic). The mean intrinsic index is relatively low, implying hard spectra. The confidence range calculated from the likelihood analysis is similar for the two distributions ($\sigma_{\rm obs}=0.34$ versus $\sigma_{\rm int}=0.35$).

\begin{figure}
\includegraphics[width=\columnwidth]{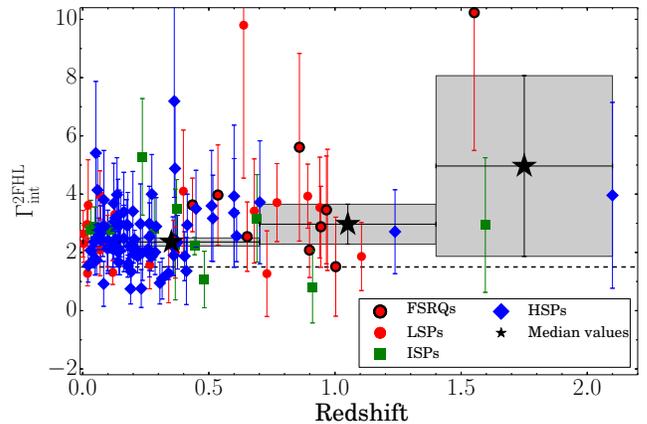}
\caption{The intrinsic spectral index versus redshift of our sample. We note that only emission mechanisms and attenuation at the sources contributes to this correlation, since the EBL effect is removed. The sources are divided between FSRQs (red circles with black edges), LSP blazars (red circles), ISP blazars (green squares), and HSP blazars (blue diamonds). Also, the median values (black stars) and their $1\sigma$ uncertainties are shown. The uncertainty of the median is estimated from the distributions in each redshift bin by bootstrapping, accounting for the uncertainties of the individual measurements. The typical lower limit of 1.5 is shown with a horizontal line.}
\label{fig2}
\end{figure}

\citet{2FHL} show in their Figure~12 that the \emph{observed} indices evolve with redshift. Alternatively, in our Figure~\ref{fig2}, we see no evidence for evolution with redshift of the \emph{intrinsic} indices in our sample, which is dominated by HSP blazars. Thus, we see no significant evolution with redshift of the emission mechanisms in this type of blazars. This conclusion needs to be interpreted carefully because of potential bias in our sample due to selection effects. In fact, at the higher redshifts we are only sensitive to the most luminous blazar populations. We also see in Figure~\ref{fig2} that most of the sources (98\%) are compatible within $1\sigma$ with having an index larger than 1.5. This 1.5 is a fiducial value typically used as a lower-limit index to derive upper limits on the EBL spectral intensity based on local observations of blazars (\ie no EBL attenuation) and theoretical arguments of blazar photon emission \citep[\eg][]{aharonian06,meyer12}. However, three sources have extremely hard spectra (low index), more than $1\sigma$ less than 1.5. From Monte Carlo simulations, we find that these 3 sources are statistically expected from our distribution of indices. These sources are listed in Table~\ref{tab1}. Two of these sources, 1ES~0502+675 \citep{benbow11} and RBS~0413 \citep{rbs0413}, have already been detected by VERITAS.

\begin{deluxetable*}{lcccc}
\setlength{\tabcolsep}{0.04in}
\tablewidth{0pt}
\tabletypesize{\scriptsize}
\tablecaption{Sources with extremely hard spectrum, $\Gamma+\Delta\Gamma\lesssim 1.5$ \label{tab1}}
\tablehead{
\colhead{2FHL Source Name} & 
\colhead{Association} &
\colhead{Redshift} &
\colhead{$\Gamma^{\rm 2FHL}_{\rm int}\pm \Delta \Gamma^{\rm 2FHL}_{\rm int}$} &
\colhead{IACT detected}
}
\startdata
2FHL J0238.4$-$3116 & 1RXS J023832.6$-$311    & 0.232 & $0.76_{-0.62}^{+0.68} $ & No\\
2FHL J0319.7+1849   & RBS~0413                & 0.19  & $0.75_{-0.52}^{+0.53} $ & \citet[][VERITAS]{rbs0413}\\
2FHL J0507.9+6737   & 1ES~0502+675            & 0.34  & $1.29_{-0.23}^{+0.22} $ & \citet[][VERITAS]{benbow11}\\
\enddata
\end{deluxetable*}

We also search in our sample for correlation between the intrinsic index and the synchrotron peak frequency as found in the 3LAC catalog \citep{3LAC}. However, no such correlation is found in our sample. The primary reason is that our sample is largely biased toward HSP blazars; therefore we are indeed missing a blazar population that is found at lower energies. Second, the spectral index scatter is larger at VHE than at HE. This second effect is the result of poor photon statistics together with the fact that at VHE we are sampling the decreasing side of the higher-energy peak of the broadband spectral energy distribution.


\begin{figure*}
\centering
\includegraphics[scale=0.28]{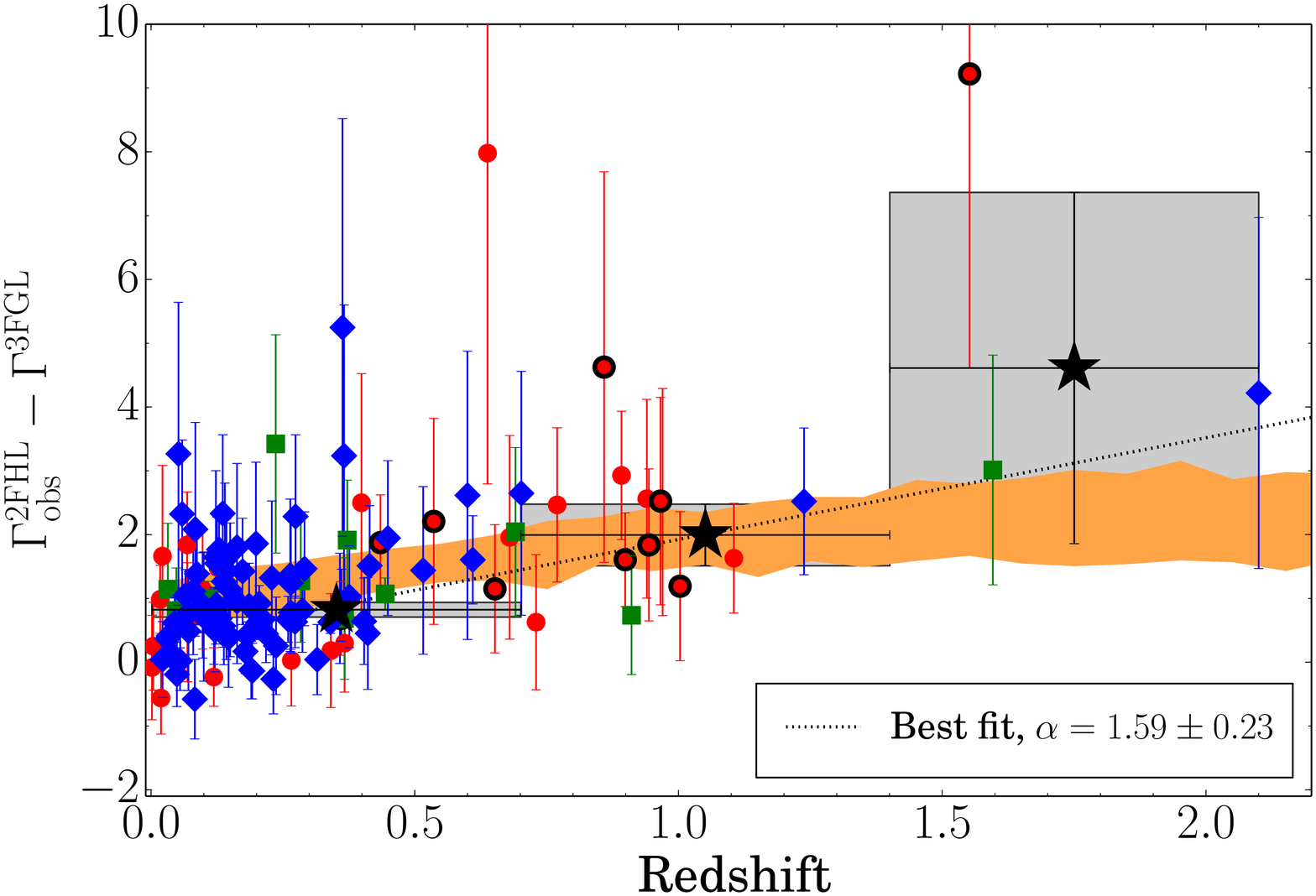}
\includegraphics[scale=0.28]{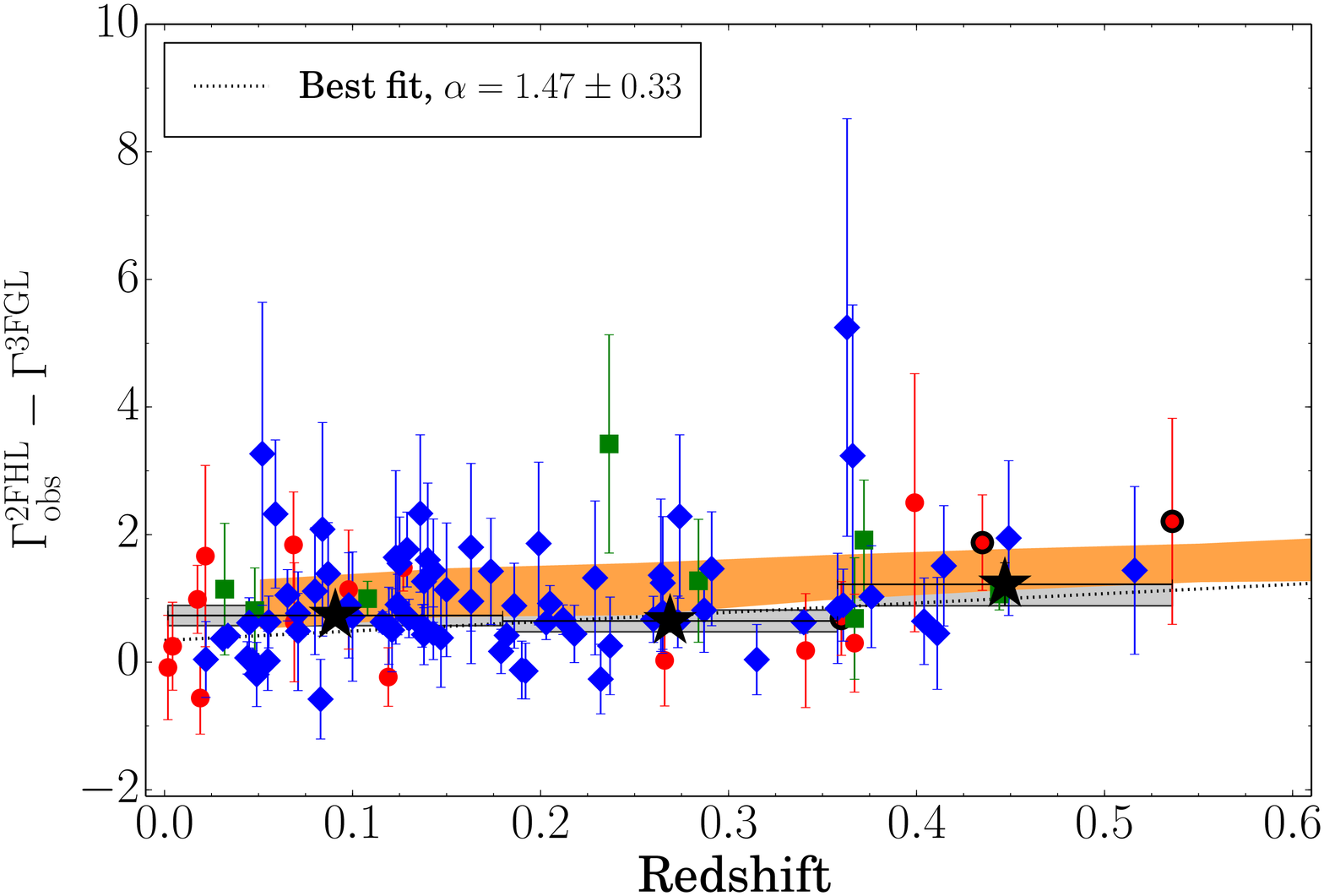}
\caption{The observed spectral break versus redshift for the full sample ({\it left panel}) and for the subsample with $z\leq 0.6$ ({\it right panel}). The predictions from simulations that include intrinsic curvature and EBL attenuation are shown with an orange band. This band encloses the $1\sigma$ uncertainties around the median of the distributions of simulated indexes in redshift bins of 0.1. The methodology to estimate the uncertainty boxes, symbols, and colors are the same as in Figure~\ref{fig2}.} 
\label{fig4}
\end{figure*}

\section{Spectral breaks}
\label{sec:results}
There are 122 sources with measured redshift in the 2FHL catalog that are also found in the 3FGL catalog \citep{3FGL}. For these sources, we complement our data with the observed HE spectral index provided in the 3FGL, $\Gamma^{\rm 3FGL}$ (column named \texttt{PowerLaw\_Index}). The comparison between the observed and intrinsic VHE indices, and the HE indices allows us to study the observed as well as the intrinsic spectral breaks. We note that previous measurements in the literature of the observed spectral break could be affected by several problems, mainly produced by the fact that those measurements compare indices from data taken by IACTs and the {\it Fermi}-LAT. Some of these potential problems are non simultaneity and dis-uniformity in the instrumental calibration and energy range for the VHE analysis, among others \citep{costamante13}. Our analysis alleviates these problems since both of our spectral indices are measured by integrating the exposure over long periods of time (48 months for the 3FGL indices and 80 months for the 2FHL indices), in the same energy bands for all sources, and with the same instrument and analysis pipeline.

\begin{figure*}
\includegraphics[scale=0.28]{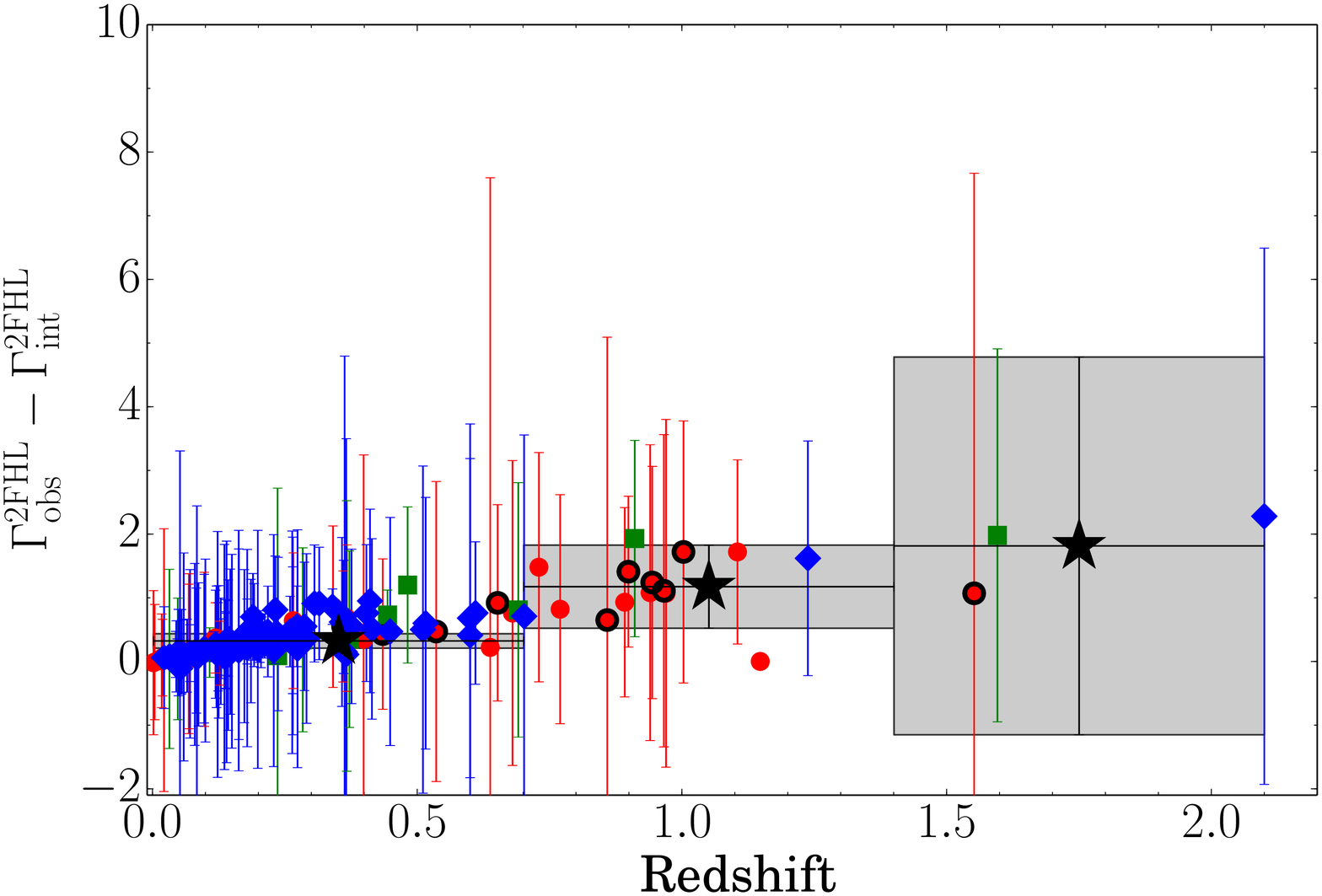}
\includegraphics[scale=0.28]{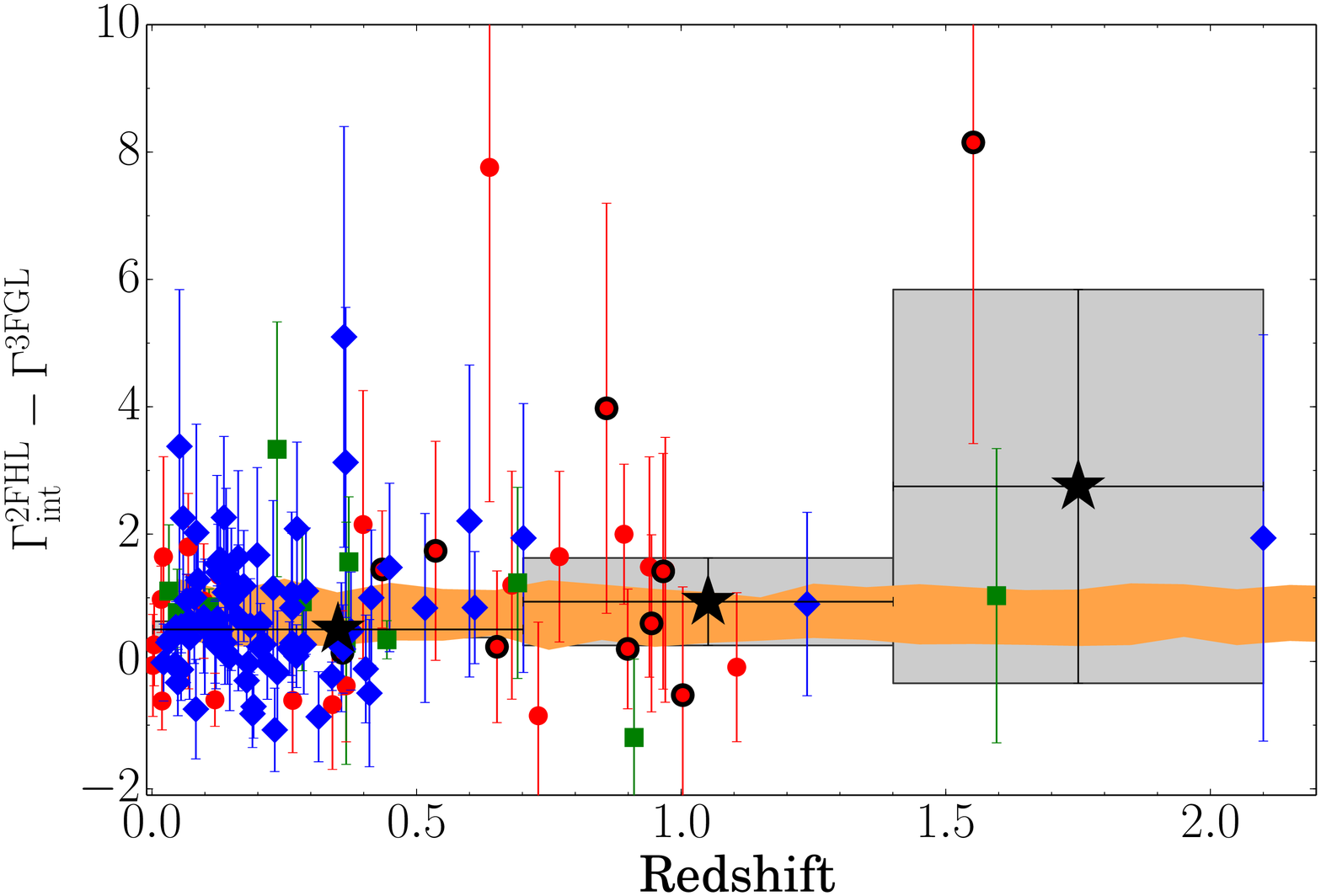}
\caption{({\it Left panel}) The difference between the observed and intrinsic indices at VHEs versus redshift. This plot shows the effect of the EBL attenuation over redshift. There is no blazar physics involved in this evolution. We note the small scatter of the index difference with redshift. There is one source (2FHL J0811.6+0146, $z=1.148$) whose indices and their uncertainties are not well estimated in the fitting since all its photons have energies slightly above 50~GeV. ({\it Right panel}) The intrinsic spectral break versus redshift. This relation removes the EBL effect and shows only the spectral effects due to the blazar physics. The methodology to estimate the uncertainty boxes, symbols, and colors are the same as in Figure~\ref{fig2}.}
\label{fig5}
\end{figure*}

Figure~\ref{fig4} shows the evolution with redshift of the observed spectral break (\ie $\Gamma^{\rm 2FHL}_{\rm obs}-\Gamma^{\rm 3FGL}= \alpha z+\beta$). The total uncertainties are calculated by adding the uncertainties from both indices. The best-fit straight lines are $(1.59\pm 0.23)z+(0.33\pm 0.05)$ for the whole sample and $(1.47\pm 0.33)z+(0.34\pm 0.05)$ for the sample of 99 sources at $z\leq 0.6$, which suffers less selection effects (that are described above). The parameter $\beta$, which is found to be larger than zero (approximately 0.3), is related to the average intrinsic blazar curvature. We note that the scatter of the LSP population is approximately double of that of the HSP population.

To study this effect, we produce simulations that include 1,000 physically motivated spectra of blazars that reproduce the range of curvature and indices of \lat blazars \citep{tramacere11}. We also include the effect of the EBL absorption \citep{dominguez11a}. Then, observations of comparable exposures as those of the 3FGL and 2FHL are generated from the model spectra and analyzed in the same way as the real data. A similar approach is used by \citet{ebl12} to validate their analysis. The results from these simulations and analysis are shown in Figure~\ref{fig4}. From the agreement between the simulations and the data, we conclude that EBL attenuation solely explains the observations. Therefore, there is no need for alternative explanations for the $\gamma$-ray photon propagation through the Universe. Examples of these alternative scenarios are the secondary cascade hypothesis of \citet{essey10,essey12} or axion-like particle conversion \citep[\eg][]{sanchez-conde09,dominguez11b,horns12,galanti15}. 

We can separate the contribution from blazar physics and the EBL effect by correcting for the EBL attenuation. Current EBL models agree on their intensities within a factor better than two in the more local Universe $z<0.6$ and at the wavelengths that interact with the VHE photons discussed here. Therefore, our results will not depend on the EBL model selection, as long as one of the recent models is used \citep[\eg][]{finke10,gilmore12,helgason12,scully14,khaire15}. Figure~\ref{fig5} plots the evolution of the difference of the observed and intrinsic indices at VHEs (\ie $\Gamma^{\rm 2FHL}_{\rm obs}-\Gamma^{\rm 2FHL}_{\rm int}$), which shows the effect of the EBL attenuation alone. Figure~\ref{fig5} shows that the EBL has little effect on Fermi 2FHL spectra at low redshift, in agreement with what is found in simulations and displayed in Figure~\ref{fig4}. We also note that the EBL contributes little to the total scatter of the observed spectral break shown in Figure~\ref{fig2}. In fact, the scatter of the LSP population is similar to those of the ISP and HSP populations, which is expected because the EBL effect is not dependent on the source type.

The effect of the EBL can be removed by plotting the difference of the intrinsic index at VHE and the index at HE (\ie $\Gamma^{\rm 2FHL}_{\rm int}-\Gamma^{\rm 3FGL}$), which we call the intrinsic spectral break. The intrinsic spectral break is shown in Figure~\ref{fig5}, where we find no dependence of its median with redshift. This indicates that our sample exhibits no evolution of the physics that drives the emission in blazars at these energies. However, we cannot conclude that this is true for each of the blazar populations since our sample includes mostly HSP blazars and at the higher redshifts only the most luminous blazars with the lowest intrinsic spectral breaks. The scatter of the intrinsic spectral break is large in our sample. In particular, the scatter is larger for the LSP and ISP populations than for the HSP blazars. This can be understood because at VHE the spectral drop of the higher energy peak of the broadband spectral energy distribution occurs more strongly in the LSP population.

\section{Summary}
\label{sec:summary}

We have presented an analysis of the intrinsic (unattenuated by the EBL) spectral indices of 128 extragalactic sources detected up to $z\sim 2$, allowing us to study the evolution of the blazar physics up to high redshifts. The median of the distribution of spectral indices is rather low (2.20) and significantly harder than the observed median index (2.54) but with a similar dispersion. There are also three extremely hard sources (approximately $1\sigma$ away from the fiducial lower limit value for the intrinsic index of 1.5) that are candidates for further study.

We also present an analysis of the evolution with redshift of the observed and intrinsic spectral breaks. The observed softening of the spectral break agrees with results from simulations that include intrinsic blazar curvature and EBL attenuation. This agreement indicates that alternative explanations for the $\gamma$-ray photon propagation such as secondary cascades or axion-like particle conversion are not necessary to explain the \lat data. Furthermore, we find that most of the scatter in the spectral break is dominated by the lower frequency synchrotron peak blazars and relates to the blazar physics rather than the EBL attenuation. We also conclude that the physical properties of blazars do not evolve with redshift in our sample, which includes mostly HSP blazars. This result supports the idea that this type of BL Lacs is a useful probe of the EBL.

\section*{Acknowledgments}
The authors thank Luca Baldini, Jean Ballet, Seth Digel, Steve Fegan, Justin Finke and Francisco Prada for helpful suggestions.

We acknowledge the support of the Spanish MICINN's Consolider-Ingenio 2010 Programme under grant MultiDark CSD2009-00064.

The \textit{Fermi}-LAT Collaboration acknowledges support for LAT development, operation and data analysis from NASA and DOE (United States), CEA/Irfu and IN2P3/CNRS (France), ASI and INFN (Italy), MEXT, KEK, and JAXA (Japan), and the K.A.~Wallenberg Foundation, the Swedish Research Council and the National Space Board (Sweden). Science analysis support in the operations phase from INAF (Italy) and CNES (France) is also gratefully acknowledged.

\bibliographystyle{apj}

\label{lastpage}
\end{document}